\documentstyle[12pt]{article}

\begin{document}

\topmargin 0pt
\oddsidemargin 5mm

\setcounter{page}{1}
\vspace{3mm}
\begin{center}
{\bf ENTANGLEMENT INDUCED RADIATION PROCESSES
WHICH ARE FIRST ORDER IN WEAK FIELD}\\
\vspace{3mm}
{\bf M.L.Ter-Mikayelyan}\\
\vspace{3mm}
{\small Institute for Physical Research, Armenian National Academy of
Sciences,378410 Ashtarak 2, Armenia, E-mail:
root@ipr.arminco.com}\\
\end{center}

\vspace{3mm}
\indent

{\bf Abstract}\\
Nonlinear processes of light scattering on a two-level system 
near resonance are considered. The problem is reduced to the 
emission and absorption of an entangled system, formed by a strong 
resonant field and a two-level system, having a non-factorizing 
wave function. Various regimes of emission and absorption of 
the entangled system are determined by the parameter 
 $\alpha =\left| 2dE/\hbar\Delta  \right| $, where
E is the amplitude of the strong resonant field, d is the
electric
dipole moment of the transition 1-2 of the unperturbed atom, 

 $\Delta $
 is the detuning off resonance, 
 $\Delta =E_{2} -E_{1} -\hbar \omega $, with
 $E_{1,2} $
 being the energies of, 
respectively, lower and upper atomic levels, and  the carrier 
frequency of the resonant field. Two limiting cases of the switching 
on the interaction between the atom and field are considered, 
those of the adiabatic and the sudden switching on; these two 
cases allow simple analytical solutions and lead to essentially 
differing physical results, either for the probabilities of the 
processes of the first order with respect to the weak field, 
or for their coherent properties. Note that at 
 $\alpha <<1$
 the conventional 
perturbation technique is applicable. This means that for weak 
fields we can use perturbation method because the corresponding 
$\alpha^\prime<<1$. The nonlinear resonance fluorescence and the amplification
of weak radiation field by the entangled system at optical frequencies 
are considered. The coherent properties of the emission of entangled 
system are studied which is important for the problems of propagation 
in gaseous media, decoherency and quantum communication.
\\Keywords; nonlinear optics, fluorescence, light amplification, entangled
system. 

\vspace{6mm}
\leftline{\bf1.Introduction}
\indent
Investigations of the optical radiation scattering by single 
atoms or in macroscopic gaseous media have almost centennial 
history. These investigations became their further development 
recently in conjunction with atom cooling and appearance of new, 
highly effective, lasers. Of special interest is the use of these 
processes in the problems of creation of one-atom lasers (see, 
e.g., [1] and references therein), and for quantum computations 
(see, e.g., [2] and references therein). In the present work 
we consider processes of the first order, with respect to the 
weak field, on a coupled system  "atom+strong resonant field"
 which will be termed below as "entangled system"
 (ES). A special attention will be paid to coherent properties 
of the emission of ES and to the ways of switching on the interaction 
between the two-level system and the strong resonant field. The 
adiabatic and sudden switching on for which the problem has simple 
solutions are the models for real situations. The parameter determining 
these two limiting cases is the ratio of the resonance detuning 
$\Delta $
to the temporal characteristics of the pump (e.g., the temporal
characteristic, 
$\tau $, of the pulse leading edge). At
$\left| \Delta \tau \right| >>1$
the adiabatic
switching is realized while at 
$\left| \Delta \tau \right| <<1$
the switching is sudden. In the
first case the atomic electrons are perturbed slightly what leads 
to the well-known Kramers-Heisenberg formula for the dispersion 
and Raman-effect [3-6]; in the second case the atomic electrons 
oscillate, by stimulated emission and absorption of photons of 
the external resonant field, between the atomic levels (see, 
e.g., [7] where is a mistake corrected in [8]). In the following 
calculations we do not introduce phenomenologically the damping, 
$\Gamma $
, of the upper level. The reasons for this are following. First, 
the time of spontaneous electric-dipole transitions is of the 
order of $10^{-8}$s, i.e., at shorter interaction times there is no
reason to introduce these transitions; second, it is assumed 
that  $\left| \Delta \right| >\Gamma $
; third, to calculate the spontaneous processes in an ES 
we have to use an other technique than that used by V. Weisskopf, 
[3-6]; finally, the frequency of population oscillations in 
a strong resonant field which is of order 
 $\left| dE/\hbar\right| $
 and plays the dominant 
role in the calculations below should be greater than the natural 
linewidth $\Gamma $
. To the best of our knowledge the use of ES in calculations 
of nonlinear resonance fluorescence and of absorption of a weak 
external radiation has been suggested in Ref.[9] where the adiabatic 
switching was studied (when comparing the present results with 
those of [9] it should be taken into account that the notations 
for the roots of the characteristic equation are different). 
For clarifying the questions of light amplification in scattering 
by an atomic system (or, in other words, of interference of radiation 
of ES formed by atoms in gaseous media or in ion traps) the study 
of the coherence properties of individual components of the scattered 
radiation is needed. These problems are closely related to those 
of information transfer and recording by means of atomic systems. 
In this conjunction we employ here the semiclassical theory of 
radiation where the phase and amplitude of strong resonant field 
may be treated simultaneously. To study the wave pictures of 
radiation processes it can be done [see, for example, [3, 6, 
8]. Using, however, the correspondence principle makes it possible 
to interpret the obtained physical results quantumelectrodynamically 
as well. Naturally, the method of correspondence [3-8], giving 
the results equivalent to quantumelectrodynamical calculations, 
is used for emission and absorption of the photons of a weak 
radiation. The generalization of the obtained results for the 
non-classical description of the strong resonant field, when 
the particles properties of light are important, requires a special 
investigation.\\
\indent
For adiabatic switching on  the interaction with a two-level
atom the wave functions of an ES will be denoted below by 
$\Phi _{1,2} $ while
for sudden switching on the interaction by 
$\Phi^ \prime _{1,2} $. As each of these
function systems is a complete orthonormilized system describing 
ES, the formulas connecting these two systems have the following 
form (see, e.g. [8]).
\begin{equation}
\label{AA}
\Phi^ \prime _{1} =C_{1} \Phi _{1} +C_{2}^{*} \Phi _{2} ,
\Phi^ \prime _{2} =-C_{2} \Phi _{1} +C_{1} \Phi _{2}
\end{equation}
\\where
\begin{equation}
\label{AB}
C_{1} =\frac{1}{\sqrt{2} } \left( 1+\frac{\Delta }{\Omega } \right)
^{1/2},
C_{2} =-\frac{\lambda _{1} }{V} C_{1}.
\end{equation}
The quantities entering $C_{1}$ and $C_{2}$ are equal to
\begin{equation}
\label{AC}
\Delta =E_{21} -\omega ,
\Omega =\sqrt{\Delta ^{2} +4\left| V\right| ^{2} } ,
\lambda _{1} =\frac{\Delta -\Omega }{2} ,
V=-\frac{Ed_{21} }{\hbar } ,
\end{equation}
with E being the amplitude of the field acting on a two-level
atom near resonance, i.e. under the condition 
\begin{equation}
\label{AD}
\left| \Delta \right| <<E_{21} ,
\end{equation}
\\where $E_{21} =E_{2} -E_{1} $
is the difference of the upper and lower levels energies
measured in $sec^{-1} (see, Fig.1)$
\begin{equation}
\label{AE}
E(r,t)=Ee^{ikr-i\omega t} +c.c.
\end{equation}
\indent
The electric dipole moment of the unperturbed two-level atom
is denoted by $d_{21}$
\begin{equation}
\label{AF}
d_{21} =d_{12}^{*} =\int \psi _{2}^{*} er\psi _{1} dV=e^{i(\varphi _{1}
-\varphi _{2} )}d,
\end{equation}
where
\begin{equation}
\label{AG}
\psi _{1,2} =u_{1,2} e^{-iE_{1,2} t}
\end{equation}
are the wave functions of a free atom in the lower and the upper 
states. Note that for our convenience in formula (6) we separated 
random phases of the atom as the wave function in quantum mechanics 
is determined to within a random phase. These phases are different 
for each single atom and denoted by 
$\varphi _{1,2}^{i}. $
When calculating the scattering
light by an ensemble of atoms it is necessary to average over 
the random phases. With the results being independent of random 
phases, such processes are coherent; otherwise, in the emission 
processes of ES the radiation intensities from different ES will 
be simply added, i.e. such processes will be noncoherent. With 
atom being acted upon by a resonance radiation and random phase 
being different at different moments, the same arguments can 
be related to one atom. In accordance with the quasiclassical 
radiation theory the probability of spontaneous emission of the 
photon with the frequency  $\omega^ \prime $
, which is equal to the difference 
of the energies of two discrete levels,  the momentum $k^\prime$ and the
polarization $e^\prime$ into the solid angle $dO^\prime$ in the transition
$2\rightarrow 1$ willbe determined by the expression [3-6]
\begin{equation}
\label{AH}
dW_{sp} =\frac{\omega^ {\prime 3} }{2\pi \hbar c^{3} } \left|
e^{\prime*} d^{-} \right| ^{2} dO^\prime,
\end{equation}
where $d^{-}$ is the negative frequency part of the dipole moment.
For a system of atoms, with coordinates $r_{i}$ expression (8) reads
[5]
\begin{equation}
\label{AJ}
dW_{sp} =\frac{\omega^ {\prime 3} }{2\pi \hbar c^{3} } \left|
\sum\limits_{i}(e^{\prime*} d_{i}^{-} )e^{i(k-k^\prime )r_{i} }  \right|
^{2} dO^\prime .
\end{equation}
\indent
Besides spontaneous emission there exist stimulated emission
and absorption the probability of which are determined by the 
same formula (8) multiplied by $n^\prime$, with $n^\prime$
being the total numberof photons with the energy
$\omega^ \prime $, the momentum $k^\prime$ and the polarization
$e^\prime$. The quantity $n^\prime$ is expressed in terms of the
spectral-angular density  $I(k^\prime ,e^\prime )$
of stimulating radiation by the following formula
\begin{equation}
\label{AK}
n^\prime (k^\prime ,e^\prime )=\frac{8\pi ^{3} c^{2} }{\hbar
\omega^{\prime3}} I(k^\prime ,e^\prime ),
\end{equation}
where the radiation intensity is equal to 
$J=\int I(k^\prime ,e^\prime )d\omega^\prime dO^\prime  $
. In the case of stimulated 
processes the solid angle should be treated as the solid angle 
in direction distribution of the stimulated radiation. In this 
case the probability of emission and absorption will be determined 
by the following simple expressions, respectively [3-6] 
\begin{equation}
\label{AL}
dW^\prime =dW_{sp} (n^\prime +1)
\end{equation}
\begin{equation}
\label{AM}
dW^\prime =dW_{sp} n^\prime .
\end{equation}
With ES being considered instead of a free atom, all the changes
to be introduced in formula (8) are reduced to replacing the 
negative frequency part of the dipole moment of the free atom 
by the negative frequency part of the ES  dipole moment.

\vspace{6mm}
\begin{center}
{\bf{2. Coherence of "entangled systems" radiation and
comparison
of electric dipole moments of "entangled systems"
at various regimes of switching}}
\end{center}
\indent
The expressions for $D_{ik} $
in the case of adiabatic switching are given
in [9] (see, also [8]) and have the following form
\begin{equation}
\label{AN}
D_{11} =-D_{22} =-\frac{V}{\Omega } d_{12} e^{-i\omega t} +c.c.
\end{equation}
\begin{equation}
\label{AP}
D_{12} =D_{21}^{*} =-\frac{1}{2} \left( 1-\frac{\Delta }{\Omega }
\right) d_{21} e^{-i(\omega -\Omega )t-2i\varphi _{0} } +\frac{1}{2}
\left( 1+\frac{\Delta }{\Omega } \right) d_{12} e^{-i(\omega +\Omega )t},
\end{equation}
with $\varphi _{0} =\varphi +\varphi _{1}^{i} -\varphi _{2}^{i} $
being  the phase of the interaction 
\begin{equation}
\label{DG}
V=\left| V\right| e^{i\varphi _{0} -i\pi },
\end{equation}
where $\varphi$ is a constant or a slowly varying phase of the field E;
$\varphi _{1,2}^{i} $
are the above-denoted random phases of the atoms. As it is seen 
from (13), the latter cancels the random phase $d_{12}$ in the expression
(13) and remains in (14), i.e. the processes described by the 
dipole moment $D_{11}$ will be coherent ones, while radiative processes
induced by the dipole moment $D_{12}$ will be noncoherent. Besides,
there is a phase of the radiation in expression (15). This phase, 
naturally, should not be changed during the interaction. A known 
example of that type is the "forward"
stars light scattering in atmosphere which is coherent and the
noncoherent "side"
stars light scattering determining the blue color of the sky.
The photons reaching the Earth should be coherent or identical, 
i.e. to be in a phase volume of 
 $\Delta x\Delta y\Delta z\Delta p_{x} \Delta p_{y} \Delta p_{z} \approx
(2\pi \hbar)^{3} $. Having applied this relation
to a star on the whole, it is easy to be convinced of the fact 
that due to giant distances to the Earth the light radiated by 
stars and detected on the Earth is coherent (see, e.g., [10]).
\indent
The dipole moments of ES for that case of sudden switching
$D_{ik}^\prime$
are obtained from the relation (1) and the expressions (13), 
(14).

a) Consider these relations in the limit of  $\alpha<1$. Omitting
simple calculations, we have 
\begin{equation}
\label{GA}
D^\prime _{11} =-D^\prime _{22} =sign\Delta D_{11} -\left( \frac{\alpha
}{2} e^{i\varphi _{0} } D_{21} +c.c.\right)
\end{equation}
\begin{equation}
\label{AA}
D^\prime _{12} =\left( 1-\frac{\alpha ^{2} }{4} \right) D_{12}
-e^{-2i\varphi _{0} } \frac{\alpha ^{2} }{4} D_{21} +\alpha
e^{-i\varphi_{0}}D_{11}   at \Delta>0
\end{equation}
\begin{equation}
\label{GE}
D^\prime _{12} =\frac{\alpha ^{2} }{4} D_{12} -e^{-2i\varphi _{0} }
\left( 1-\frac{\alpha ^{2} }{4} \right) D_{21} +\alpha
e^{-i\varphi_0}D_{11}  at  \Delta<0 
\end{equation}
\indent
In accordance with the expression (14) at $\alpha<1$ we have
\begin{equation}
\label{AB}
D_{11} =-D_{22} =\frac{\alpha }{2} \left( 1-\frac{\alpha ^{2} }{2}
\right) e^{i\varphi _{0} -i\omega t} d_{12} +c.c.
\end{equation}
\begin{equation}
\label{AN}
D_{12} =-\frac{\alpha ^{2} }{4} d_{21} e^{i(\omega -\Omega )t-2i\varphi
_{0} } +\left( 1-\frac{\alpha ^{2} }{4} \right) d_{12}
e^{-i(\omega+\Omega)t} at  \Delta>0
\end{equation}
\begin{equation}
\label{GK}
D_{12} =-\left( 1-\frac{\alpha ^{2} }{4} \right) d_{21} e^{i(\omega
-\Omega )t-2i\varphi _{0} } +\frac{\alpha ^{2} }{4} d_{12} e
^{-i(\omega+\Omega)t} at \Delta<0
\end{equation}
\\b)In the case of $\alpha>1$ the same relations take the following
form:
\begin{equation}
\label{AC}
D^\prime _{11} =-D^\prime _{22} =\frac{sign\Delta }{\alpha } D_{11}
-\frac{1}{2} \left( e^{i\varphi _{0} } D_{12} +c.c.\right)
\end{equation}
\begin{equation}
\label{AD}
D^\prime _{12} =\frac{1}{2} \left( 1+\frac{sign\Delta }{\alpha } \right)
D_{12} -\frac{1}{2} e^{-2i\varphi _{0} } \left( 1-\frac{sign\Delta
}{\alpha } \right) D_{21} +D_{11} e^{-i\varphi _{0} }
\end{equation}
$D_{11}$ and $D_{12}$ entering (22) and (23) at $\alpha>1$ are equal to:
\begin{equation}
\label{FE}
D_{11} =-\frac{1}{2} \left( 1-\frac{1}{2\alpha ^{2} } \right) \left(
e^{-i\varphi _{0} +i\omega t} d_{21} +c.c.\right)
\end{equation}
\begin{equation}
\label{AE}
D_{12} =\frac{1}{2} \left( 1+\frac{sign\Delta }{\alpha } \right) d_{12}
e^{-i\left( \omega +\Omega \right) t} -\frac{1}{2} \left(
1-\frac{sign\Delta }{\alpha } \right) d_{21} e^{-i\left( \omega -\Omega
\right) t-2i\varphi _{0} }
\end{equation}
\indent
Expressions (16), (19), (22), (24) will determine coherent processes
while the expressions (17), (18), (20), (21),
(23), (25) noncoherentones. It appears at once  from the above that
during sudden switching
on the interaction there is a new qualitative phenomenon - noncoherent
Rayleigh scattering caused by the presence of the 
$D_{11}$ term initiating
Rayleigh scattering in the expressions (17), (18) and (23). On
the analogy of that, the term of 
 $D_{12}$ in (16) and (22) causes the
appearance of coherent radiation at the $\omega\pm\Omega$ side 
frequencies, which
is always noncoherent during adiabatic switching. During the 
interaction between the resonant radiation (depending on the 
way of switching on the interaction) and the system consisting 
of some two-level atoms as well during propagation and amplification 
through a gaseous medium it leads to new qualitative phenomena.

\vspace{6mm}
\begin{center}
{\bf{3. Comparison of the probability  of the first-order processes
at various regimes  of switching on the interaction}}
\end{center}
\indent

As it was shown in item 1 the calculation of the probability
of the resonant radiation scattering on a two-level system is 
reduced to the calculation of ES emission. For this purpose in 
the expression (8) it is necessary to change the negative component 
of the dipole moment of the free atom by the negative frequency 
part of the ES dipole moment $D_{ik}$ in the case of adiabatic switching
on the interaction and by 
 $D^\prime _{ik}$ in the case of sudden switching on
the interaction, respectively. For absorption processes it is 
necessary to substitute the positive frequency parts of the corresponding 
dipole moments of ES 
 $D_{ik}^{+}$ and $D_{ik}^{\prime+}$
 in the expression (8) and make use of the formulas (11) and
 (12) for emission and absorption, respectively. As it is seen from
the previous item the dipole moments $D_{ik}$  and
$D_{ik}^\prime$ have the components 
at the resonant and the side $(\omega\pm\Omega)$ frequencies.
Fig. 1. represents
the dependence of the side frequencies upon  for  $\Delta>0$ and
$\Delta<0$. It is obvious that the radiations at different frequencies
 do not interfere with each other and should be calculated separately. 
As to emission and adsorption at the same frequency, they can 
compensate each other in stimulating processes if 
 $D_{ik}^{+} =D_{ik}^{-} $ or
 $D^{\prime+}_{ik} =D^{\prime-} _{ik} $. As a
result of there remains only a spontaneous emission. If 
 $D_{ik}^{+} \neq D_{ik}^{-} $ or
 $D^{\prime+} _{ik} \neq D^{\prime-}_{ik}$
there is no complete compensation of stimulated processes of 
emission and absorption and to obtain a resulting radiation one 
needs to subtract one from another. The probability of absorption 
is conditionally considered to be negative. These probabilities 
are set on the negative direction of the ordinate-axis in Fig. 
2 and 3. The stated rules take place during the radiation processes 
of the first order with respect to a weak field. The radiative 
processes (termed as parametric ones considered in Ref. [11]) 
of the second order can also be considered by using  the qusiclassical 
theory of emission. They may become essential in forward scattering 
. The effect of the processes of the second-order with respect 
to a weak field is not considered in the present paper.
\\
\indent
We begin by considering the case of adiabatic switching on the
interaction. The probability of spontaneous emission at the frequency 
$\omega^\prime=\omega$ is coherent one and given by the following
expression at  $\alpha<1$  for the transition
$\Phi_{1}\rightarrow\Phi_{2}$
\begin{equation}
\label{AF}
dW_{11} =\frac{\alpha ^{2} }{4} \left( 1-\alpha ^{2} \right) dW_{sp}
(\omega )
\end{equation}
\indent
The probabilities of emission and absorption at the side frequencies
$\omega\pm\Omega$ are noncoherent ones and given by the following
formulas for the transition $\Phi_{1}\rightarrow\Phi_{2}$ at $\Delta>0$
\begin{equation}
\label{AG}
dW_{21} \left( \omega -\Omega \right) =\frac{\alpha ^{4} }{16} dW_{sp}
(\omega -\Omega )\left[ n(\omega -\Omega )+1\right]
\end{equation}
\begin{equation}
\label{GK}
dW_{21} \left( \omega +\Omega \right) =-\left( 1-\frac{\alpha ^{2} }{2}
\right) dW_{sp} (\omega +\Omega )n(\omega +\Omega ).
\end{equation}
\\Under the condition  $\Delta<0$ in the expressions (27,28) the
coefficients
and signs are changed over
\begin{equation}
\label{KA}
dW_{21} \left( \omega -\Omega \right) =\left( 1-\frac{\alpha ^{2} }{2}
\right) dW_{sp} (\omega -\Omega )\left[ n(\omega -\Omega )+1\right] 
\end{equation}
\begin{equation}
\label{KE}
dW_{21} \left( \omega +\Omega \right) =-\frac{\alpha ^{4} }{16} dW_{sp}
(\omega +\Omega )n(\omega +\Omega ).
\end{equation}
These processes are both noncoherent ones.
\\
\indent
In the case of $\alpha^{2}\gg1$ the probability of the coherent emission
of quanta with  the frequency  is coherent and given by the expression
\begin{equation}
\label{AH}
dW_{11} =\frac{1}{4} \left( 1-\frac{1}{\alpha ^{2} } \right)
dW_{sp} .
\end{equation}
\indent
The probabilities of emission and absorption at the side frequencies
 for the transition $\Phi_{1}\rightarrow\Phi_{2}$ are noncoherent and equal to
\begin{equation}
\label{AJ}
dW_{21} \left( \omega -\Omega \right) =\frac{1}{4} \left(
1-\frac{2sign\Delta }{\alpha } \right) dW_{sp} (\omega -\Omega )\left[
n(\omega -\Omega )+1\right]
\end{equation}
\begin{equation}
\label{AM}
dW_{21} \left( \omega +\Omega \right) =-\frac{1}{4} \left(
1+\frac{2sign\Delta }{\alpha } \right) dW_{sp} (\omega +\Omega )n(\omega
+\Omega ).
\end{equation}
\indent
For comparison convenience the curves for $\omega\pm\Omega$  are shown in Fig.
1 in dependence upon $\omega$ at $\Delta>0$ and $\Delta<0$. Fig. 2 shows the
probabilities of coherent and noncoherent processes for adiabatic 
switching on in terms $dW_{sp}$ depending upon the frequency at
$\alpha<1$ on the left side of Fig. 2 and on the right one at
$\alpha>1$.
The straight line above the abscissa-axis represents emission 
and that below the axis represents absorption.
\indent
\\Let us turn now to the probabilities of
emission and absorption in the case of sudden switching on the 
interaction. In this case, as it follows from the relations (1) 
 the separation of coherent and noncoherent processes is getting 
rather complicated. It is easy to see that coherent processes 
will involve the processes not only at a non-shifted frequency 
but also at the side frequencies of $\omega\pm\Omega$.
Analogous phenomena will
be also observed during noncoherent processes where besides of 
noncoherent processes for shifted frequency there appears noncoherent 
scattering at a non-shifted frequency.\\
\indent
First consider coherent processes, i.e. emission and absorption
at $\Phi^\prime_{1,2}\rightarrow\Phi^\prime_{1,2}$
when the wave function of an ES does not change. By using
the formulas (17,18) for the probabilities of the radiative
first order processes with respect to a weak field we have at 
$\alpha^{2}\ll1$:
\begin{equation}
\label{BA}
dW^\prime _{11} (\omega )=\frac{\alpha ^{2} }{4} \left( 1-\alpha ^{2}
\right) dW_{sp} (\omega )
 at  \Delta>0 and \Delta<0
\end{equation}
\begin{equation}
\label{FA}
dW^\prime _{11} (\omega -\Omega )=\frac{\alpha ^{6} }{64} dW_{sp} (\omega
-\Omega )
  at  \Delta>0 ; 
\end{equation}
\begin{equation}
\label{FB}
dW^\prime _{11} (\omega -\Omega )=\frac{\alpha ^{2} }{4} \left(
1-\frac{\alpha ^{2} }{2} \right) dW_{sp} (\omega -\Omega )
 at  \Delta>0;
\end{equation}
\begin{equation}
\label{FC}
dW^\prime _{11} (\omega +\Omega )=\frac{\alpha ^{2} }{4} \left(
1-\frac{\alpha ^{2} }{2} \right) dW_{sp} (\omega +\Omega )
 at   \Delta>0;
\end{equation}
\begin{equation}
\label{FD}
dW\prime _{11} (\omega +\Omega )=\frac{\alpha ^{6} }{64} dW_{sp} (\omega
+\Omega ) at \Delta<0.
\end{equation}

\indent
The corresponding probabilities for $\alpha^{2}\gg1$
 with taking into account the formulas (22), (24) and (25) take
the following form:
\begin{equation}
\label{FE}
dW^\prime _{11} (\omega )=\frac{1}{4\alpha ^{2} } \left(
1-\frac{1}{\alpha ^{2} } \right) dW_{sp} (\omega ),
\end{equation}
\begin{equation}
\label{FG}
dW^\prime _{11} (\omega +\Omega )=\frac{1}{16} \left( 1+\frac{2sign\Delta
}{\alpha } \right) dW_{sp} (\omega +\Omega ),
\end{equation}
\begin{equation}
dW^\prime _{11} (\omega -\Omega )=\frac{1}{16} \left( 1-\frac{2sign\Delta
}{\alpha } \right) dW_{sp} (\omega -\Omega ).
\end{equation}
\indent
Analogous expressions (34-41) take place also for the
 $\Phi^\prime _{2} \rightarrow\Phi^\prime _{2} $
 transitions 
which follows from the expressions (16) and (22).
\indent
Determine the probabilities of noncoherent processes of the first 
order with respect to the weak field by proceeding from the expressions 
(17), (18), (23).
\indent
The probability of the non-shifted scattering is independent 
of the sign of $\Delta$ and has the following form:
\begin{equation}
\label{FH}
dW^\prime _{21} (\omega )=\frac{\alpha ^{4} }{4} \left( 1-\alpha ^{2}
\right) dW_{sp} (\omega ).
\end{equation}
\indent
The probabilities of emission with the frequency $\omega-\Omega$
 and absorption for $n>\frac{\alpha ^{4} }{16} $
with the frequency $\omega+\Omega$ have the following form
at  $\Delta>0$:
\begin{equation}
\label{FJ}
dW^\prime _{21} (\omega +\Omega )=\left[ -n(\omega +\Omega )+\frac{\alpha
^{4} }{16} \right] dW_{sp} (\omega +\Omega )
\end{equation}
\begin{equation}
dW^\prime _{21} (\omega -\Omega )=\frac{\alpha ^{4} }{16} \left[ n(\omega
-\Omega )+1\right] dW_{sp} (\omega -\Omega ).
\end{equation}
\indent
The probabilities of emission of a quantum
with the frequency $\omega+\Omega$ and absorption for
 $n>\frac{\alpha ^{4} }{16} $
of a photon with the frequency $\omega-\Omega$ have the following
form at  $\Delta<0$:
\begin{equation}
\label{FK}
dW^\prime _{21} (\omega +\Omega )=\frac{\alpha ^{4} }{16} \left[ n(\omega
)+1\right] dW_{sp} (\omega +\Omega )
\end{equation}
\begin{equation}
\label{FL}
dW^\prime _{21} (\omega -\Omega )=\left[ -n(\omega -\Omega )+\frac{\alpha
^{4} }{16} \right] dW_{sp} (\omega -\Omega ).
\end{equation}
\indent
The probabilities of the noncoherent processes of the first-order
with respect to the weak field at the 
 $\Phi^ \prime _{1} \rightarrow \Phi^\prime _{2} $
 transition with taking 
into account the formulas (23) - (25) for $\alpha^{2}\gg1$ have the
following
form:
\begin{equation}
\label{FN}
dW^\prime _{21} (\omega )=\frac{1}{4} \left( 1-\frac{1}{\alpha ^{2} }
\right) dW_{sp} (\omega )
at  \Delta>0 and  \Delta<0
\end{equation}
\begin{equation}
\label{FM}
dW^\prime _{21} (\omega +\Omega )=\left[ -\left( \frac{sign\Delta
}{4\alpha } +\frac{1}{2\alpha ^{2} } \right) n(\omega +\Omega
)+\frac{1}{16} \left( 1-\frac{2}{\alpha ^{2} } \right) \right] dW_{sp}
(\omega +\Omega )
\end{equation}
\begin{equation}
\label{MF}
dW^\prime _{21} (\omega -\Omega )=\left[ \left( \frac{sign\Delta
}{4\alpha } -\frac{1}{2\alpha ^{2} } \right) n(\omega -\Omega
)+\frac{1}{16} \left( 1-\frac{2}{\alpha ^{2} } \right) \right] dW_{sp}
(\omega -\Omega ).
\end{equation}
\indent
So long as there exist the relation
 $D^\prime _{12} =D^{\prime*}_{12}$
 it is easy to obtain corresponding 
expressions for the first order radiation processes when at t=0 
wave function of ES is equal to $\psi_{2}$.\\
\indent
The dependencies of the probabilities of coherent and noncoherent
processes during sudden switching are presented in Fig. 3; for 
 $\alpha<1$ on the left and for $\alpha>0$ on the right.\\
\indent
It is easy to see that the maximum of the processes of the first
order with respect to a weak field occurs at the value of    
$\alpha\approx1$.
\indent
Spontaneous processes of emission are termed as resonance fluorescence
and in linear regime were considered in 30s [3-6]. In nonlinear 
regime they became to be under consideration  in 50s due to the 
development of laser technique. In one of the last experimental 
work provided by H. Walther's laboratory (see [12] and the references 
on the previous publications therein) they succeeded in measuring 
the bandwidth of fluorescence line to phenomenal accuracy (about 
few Hz) by using the heterodyne method for frequency measuring 
[12]. The processes of a probe field absorption were begun to 
be studied due to laser creation in 50s (see, related reference 
in [8]). The process of the side line amplification in  the optical 
range at the $\omega-\Omega$ frequency, called by authors three-photon
radiation,and $\alpha\sim1$ in the experiments on propagation
of resonance radiation
through a gaseous medium was discovered for the first time  and 
investigated by Movsesyan M. et al in 1968 [13]. The detailed 
investigation in this field on the atoms of Na and comparison 
with a theory of B.R. Mollow [14] was performed in [15].

\vspace{6mm}
\leftline{\bf{4. Conclusion}}

As it has already been noted the formulas given in the paper
can be of great use in various fields of atomic physics and quantum 
theory of information recording, storage and transfer. It is 
desirable to carry out more detailed research using quantum theory 
of emission that in some cases can result in new phenomena. The 
expressions given in this paper for probabilities of first-order 
emission with respect to a weak field, photons of that are different 
from those of strong resonant field, allow to calculate the width 
of ES quasienergy levels.\\
\indent
If we define the level width of unperturbed atom by
 $\Gamma=\int dW _{sp} $
, then the 
width of ES which is in $\Phi_{1}$ state will be determined by the probabilities
(26)-(32). If $n(k^\prime,e^\prime) = 0$, then in the case of adiabatic
switching the width of wave function $\Phi_{1}$ will be equal to
\begin{equation}
\label{LA}
\Gamma(ES)=\frac{\sqrt{1+\alpha ^{2} } -sign\Delta }{2\sqrt{1+\alpha ^{2} } }
\Gamma=n_{2}\Gamma ,
\end{equation}
where $n_{2}$ is a weight of $\psi_{2}$ state in $\Phi_{1}$
function, i.e. it determines
the probability of electron being in the upper level of unperturbed 
atom. The decay of ES occurs owing to spontaneous decay of unperturbed 
atom upper level. At $\alpha_{2}\gg1$ the $\Phi_{1}$ state will be decaying
with the width equal to $\Gamma(ES)=\Gamma/2$
 as the electron spends half of time in the upper level, i.e.
in $\psi_{2}$.
\\
\indent
At $\alpha\rightarrow0$ $\Phi_{1}\rightarrow\psi_{1}$
(for adiabatic switching at  $\Delta>0$) and naturally in
this case $\Gamma(ES)=0$
. However, at the presence of weak photons different from
photons of resonant pumping either in direction or in polarization 
there will also arise stimulated widths which as spontaneous 
ones are dependent on the state in which ES is.
\\
\indent
The decoherence problem of ES has attracted great attention of 
specialists working in the field of quantum memory and information 
transfer. These problems can be investigated for our case with 
the decay probabilities given in this paper.\\

\vspace{10MM}
\centerline{\bf{ACKNOWLEDGMENTS}}
\indent
This investigation has been supported by centralized state sources 
of the Republic of Armenia in the frame of theme 96-772 and under 
grant 170ep of J.S.S.E.P.\\
\indent
I would like to thank Prof. H. Walther and M. Loeffler for discussions
and hospitality during my visit to Garching.

\newpage
\centerline{\bf{Figure captions}}
\indent
Fig. 1.  The dependence of $\omega+\Omega$ (upper curve) and
$\omega-\Omega$ (lower curve)
frequencies on  is presented in Fig. 1.  The curves
$\omega\pm\Omega$ are symmetric
with respect to $E_{21}=\omega$ line (dash-dot). At
$\alpha^{2}\rightarrow0$ the $\omega-\Omega$ line at $\Delta<0$
is approaching to abscissa axis and at  $\Delta>0$ to
$2\omega-E_{21}$ (dot).
At $\alpha^{2}\ll1,  \omega-\Omega=\omega
-\left|\Delta\right|(1+\alpha^{2})$,
 $\omega +\Omega =\omega +\left|\Delta\right|(1+\alpha^{2}).$
 At    $\alpha^{2}\gg1$, $\omega-\Omega=\omega-\left|2V\right|-
 \frac{\Delta }{2\alpha }$;
$\omega+\Omega=\omega+\left|2V\right|+\frac{\Delta}{2\alpha}$.
\indent
The scheme of two-level atom interacting with resonant field 
at  $\Delta>0$  and  $\Delta<0$  is illustrated in the upper part on the
left.\\
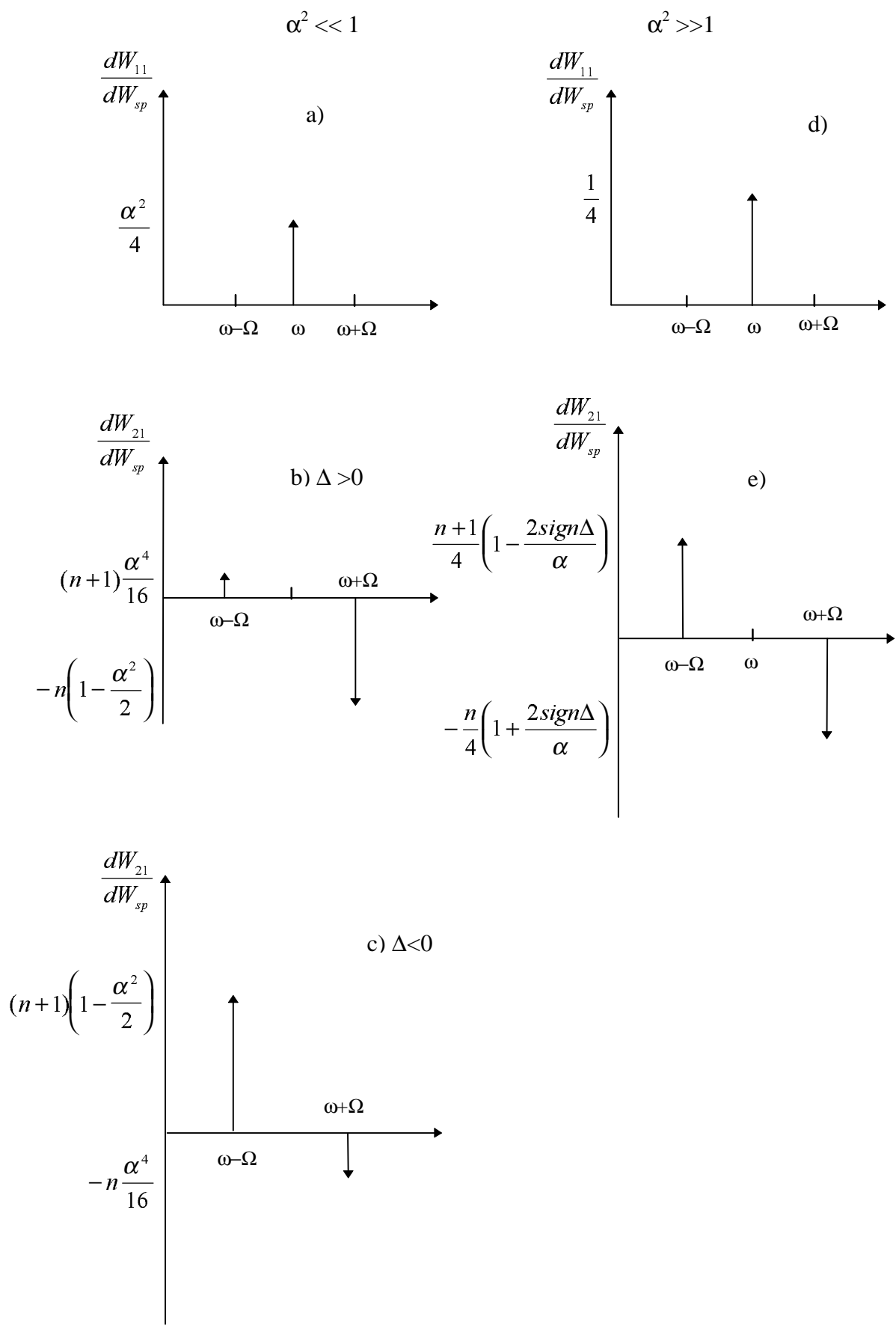
Fig. 2.  The graphs for radiative processes probabilities at
adiabatic switching on the interaction are presented. Frequencies 
are set on abscissa-axis.  The probabilities of emission (above 
abscissa-axis) and absorption (below abscissa-axis) are set on 
ordinate-axis. The probabilities are given in units of spontaneous 
probabilities at corresponding frequencies. The left column of 
graphs corresponds to the parameter of  $\alpha^{2}\ll1$ ; the right
one to  $\alpha^{2}\gg1$. The upper graphs a) and d) correspond to coherent
processes at which the wave function of ES does not change in 
the processes of emission and absorption of weak probe field. 
The graphs b), c) and e) correspond to the probabilities of noncoherent 
processes at $\Phi_{1}\rightarrow\Phi_{2}$ transition.
To obtain corresponding probabilities
for resonant fluorescence (spontaneous processes) the probabilities 
for spontaneous parts from the graphs b), c) and e) should be 
added to the graphs a) and d) at $\alpha^{2}\gg0$  and
$\alpha^{2}\ll0$ correspondingly.
The graphs have an illustrative character.
Fig. 3.  The graphs for radiative processes probabilities for
sudden switching on the interaction are presented. Frequencies 
are set on abscissa-axis. The probabilities of emission (above 
abscissa-axis) and absorption (below abscissa-axis) are set on 
ordinate-axis. The probabilities are given in units of spontaneous 
probabilities at corresponding frequencies. The left column of 
graphs corresponds to the parameter of  $\alpha<1$, the right one
to  $\alpha>1$. The upper graphs a), b) and d) correspond to coherent
processes at which the wave function of ES does not change in 
the processes of emission and absorption of weak probe field. 
The graphs c), e) and f) correspond to the
$\Phi_{1}^\prime\rightarrow\Phi_{2}^\prime$ processes. It
is accepted in the graph c) that 
 $n>\frac{\alpha ^{4} }{16} $
 and  $\Delta>0$. In the case of
 $\Delta<0$ and
 $n>\frac{\alpha ^{4} }{16} $
 the directions of vertical lines and their values 
should be changed over, i.e. the quantum with $\omega-\Omega$ frequency will
be absorbed in accordance with formula $(31^\prime)$; the quantum with
$\omega+\Omega$ frequency will be emitted according to formula
$(30^\prime)$. These lines
are dotted ones in graphic c). It is accepted on graph e) and 
f) that 
 $\frac{1}{16} >\frac{n}{4\alpha } $
. The relative height of vertical lines on graph e) corresponds 
to the conditions  $\Delta>0$ and on graphic f) to  $\Delta<0$ . If the
opposite inequality is satisfied i.e. 
 $\frac{1}{16} <\frac{n}{4\alpha } $
,  then the vertical line 
at frequency $\omega-\Omega$ at $\Delta<0$ on graph f)
will be below abscissa-axis
and the vertical line on graph e) for $\omega+\Omega$
frequency for  $\Delta>0$  will
be below abscissa-axis. 
\indent
To obtain the corresponding probabilities for resonant fluorescence 
(spontaneous processes) the probabilities for spontaneous parts 
of noncoherent processes should be added to the probabilities 
on graphs a) and d). The graphs have illustrative character.
\indent
It is followed from the figure that at  $\alpha^{2}\gg1$ that Rayleigh
scattering is completely noncoherent. The probabilities of spontaneous 
emissions at side frequencies consist of coherent and noncoherent 
components equal to each other (compare with figures in [5], 
where figure
b
 is not correct).
Fig. 4 represents spontaneous induced emission processes of
entangled system
 (resonance fluorescence). The graphs a), b), and e) correspond 
to sudden adiabatic resonance. The graphs c), d) and f) correspond 
to adiabatic one. 
Fig. 5 shows absorption and emission of a probe field by
entangled system
 for sudden and adiabatic switching. The graphs a), b), e) and f) 
correspond to sudden switching. The graphs c), d) and g) correspond 
to adiabatic one.

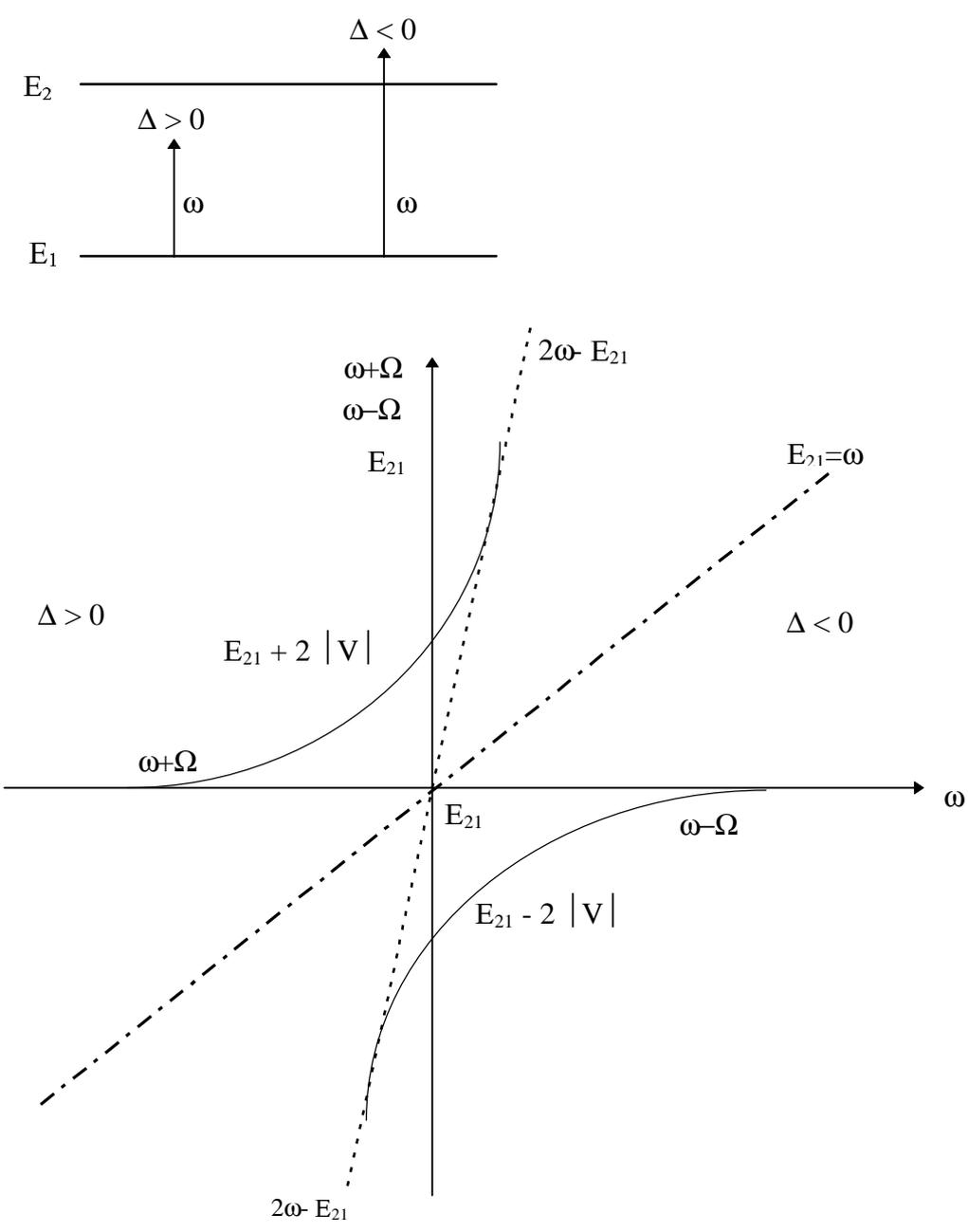
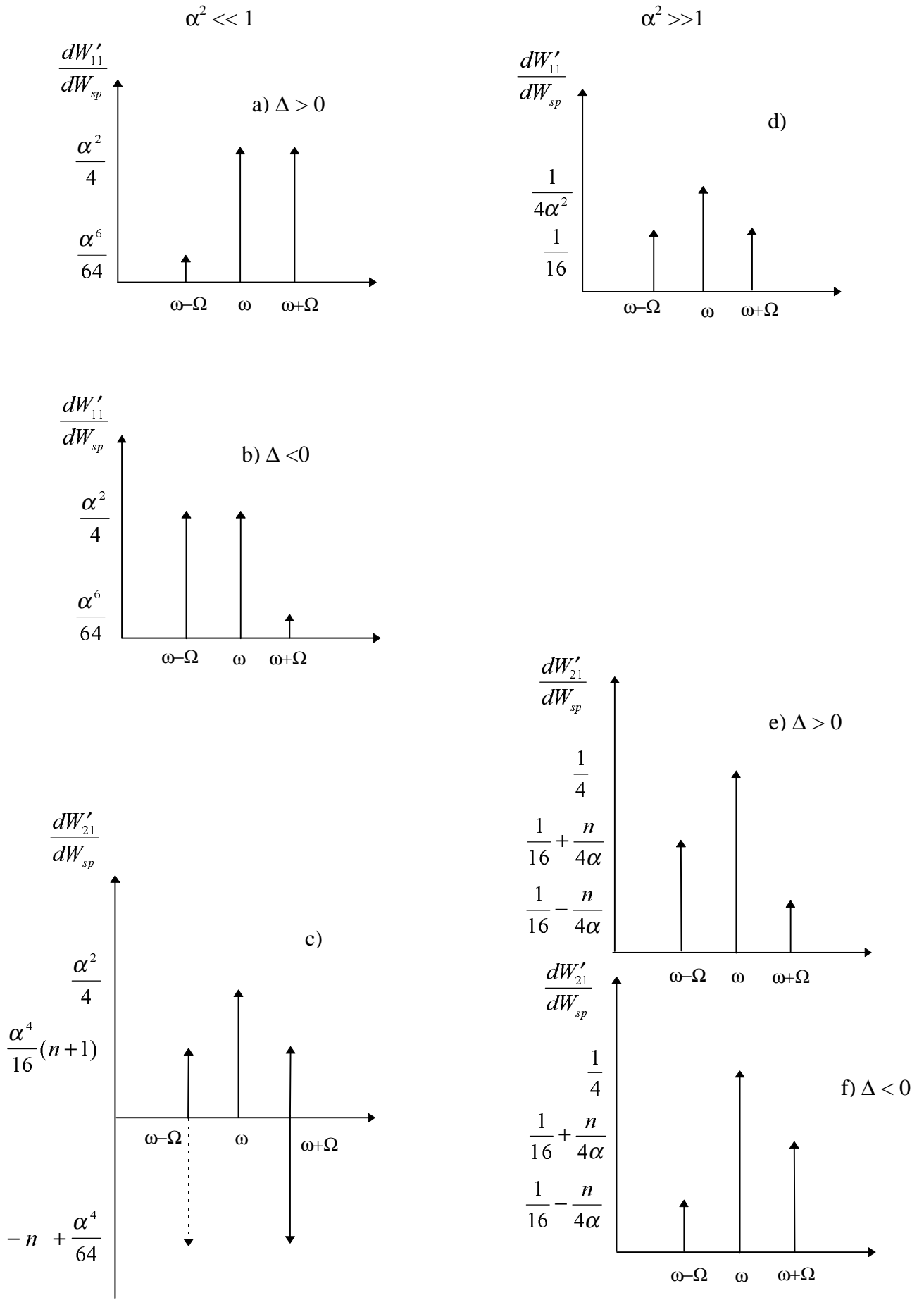
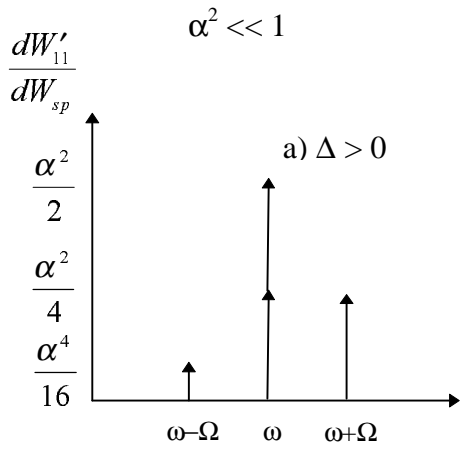
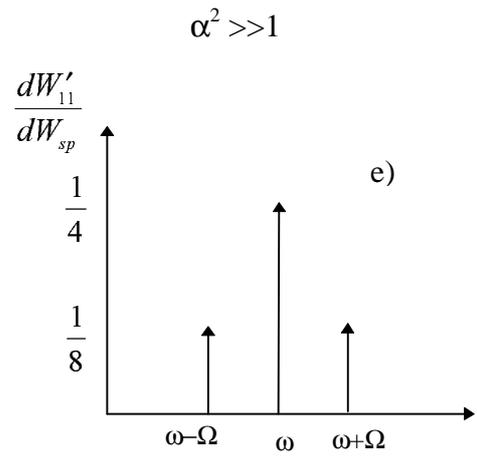
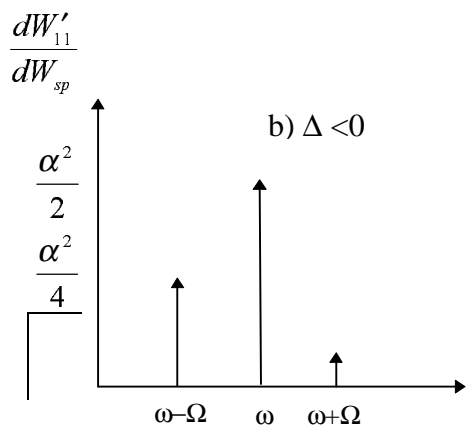
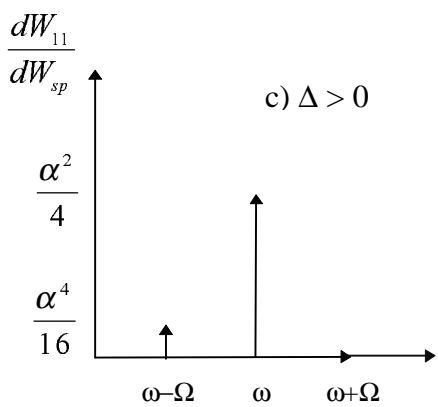
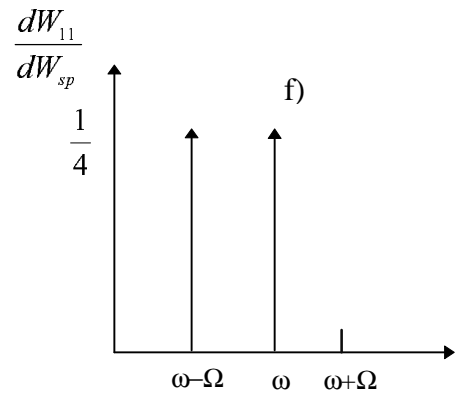
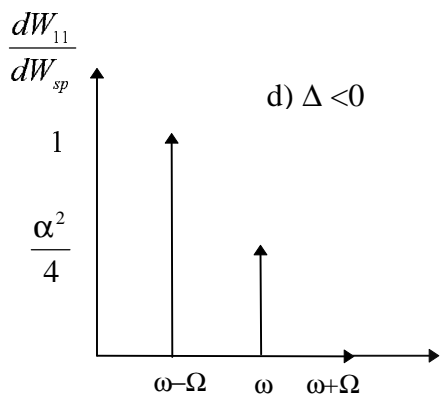
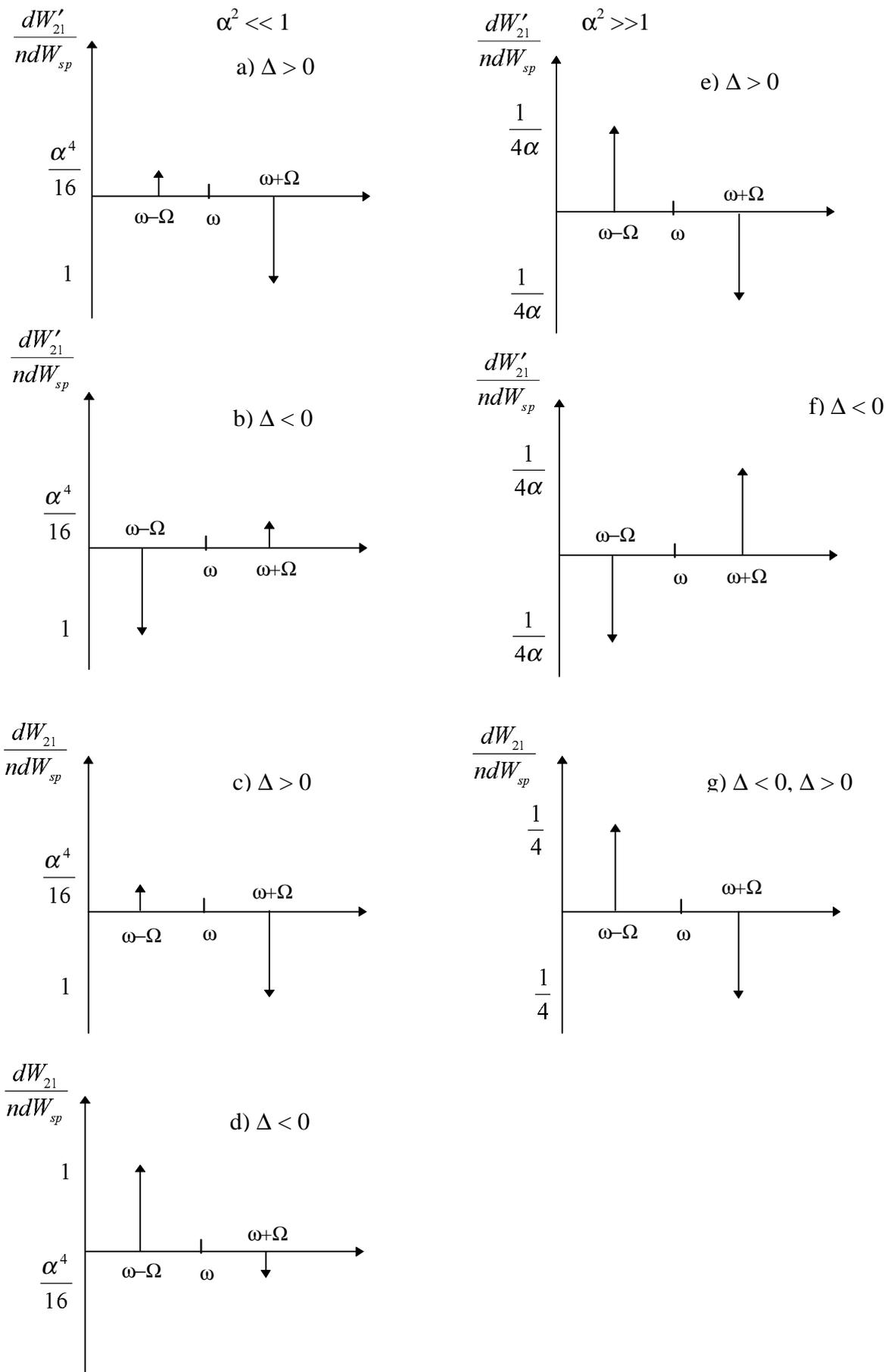
\end{document}